
\input harvmac

\def\O{{\cal O}}
\def\ne{{\nu_e}}
\def\nm{{\nu_\mu}}
\def\nt{{\nu_\tau}}
\def\l{\lambda}
\def\H{{\cal H}}
\def\bl{{\bar\ell}}
\def\mod{{\rm mod}}
\def\tM{\tilde M}

\def\tV{{\tilde V}}
\def\ra{\rightarrow}
\def\SP{S^\prime}
\def\gsim{\ \rlap{\raise 2pt \hbox{$>$}}{\lower 2pt \hbox{$\sim$}}\ }
\def\lsim{\ \rlap{\raise 2pt \hbox{$<$}}{\lower 2pt \hbox{$\sim$}}\ }

\Title{hep-ph/9502418, WIS-95/7/Mar-PH}
{\vbox{\centerline{Lepton Mass Matrix Models}}}
\bigskip
\centerline{Yuval Grossman and Yosef Nir}
\smallskip
\centerline{\it Department of Particle Physics}
\centerline{\it Weizmann Institute of Science, Rehovot 76100, Israel}
\bigskip
\baselineskip 18pt

\noindent
The smallness and hierarchy in fermion parameters
could be the result of selection rules due to
an Abelian horizontal symmetry broken by a small parameter.
When applied to the lepton sector, then for a large class of models,
a number of interesting order of magnitude relations arise: with
$i<j$, $m(\nu_i)/m(\nu_j)\sim\sin^2\theta_{ij}$;
$m(\ell^-_i)/m(\ell^-_j)\lsim\sin\theta_{ij}$;
$m(\nu_i)/m(\nu_j)\gsim m^2(\ell^-_i)/m^2(\ell^-_j)$;
$m(\ne)\lsim m(\nm)\lsim m(\nt)$. The relations
between neutrino masses and mixings may become exact if the
horizontal symmetry together with holomorphy induce certain zero
entries in the lepton mass matrices. A full high energy theory is likely
to include scalars with flavor changing couplings and heavy leptons in
vector representations; however, the masses of these particles are
too heavy to be directly observed in experiment. Indirect evidence
for the horizontal symmetry may arise from other sectors of the theory:
non-degenerate sleptons are allowed as the symmetry aligns lepton and
slepton mass matrices; light leptoquarks are allowed as the symmetry
can make their couplings diagonal and chiral.

\Date{}

\newsec{Introduction}
Only three parameters of the lepton sector are unambiguously
experimentally determined. These are the three charged lepton masses
\ref\PDG{Particle Data Group, Phys. Rev. D50 (1994) 1173.}:
\eqn\chargedmasses{
m_e=0.5110\ MeV,\ \ \
m_\mu=105.7\ MeV,\ \ \
m_\tau=1777\ MeV.}
The unambiguous information on neutrino masses consists of only
upper bounds \PDG:
\eqn\numasses{
m_\ne\leq5.1\ eV,\ \ \
m_\nm\leq0.16\ MeV,\ \ \
m_\nt\leq31\ MeV.}
However, various theoretical arguments and observational puzzles
suggest that neutrinos may have non-vanishing masses. For example,
if $\nt$ is to be an important component in the dark matter then
\eqn\darkmatter{
m_\nt=\O(10\ eV).}
If the solar neutrino problem is solved by the MSW mechanism
with $\ne-\nm$ oscillations then
\ref\HaLa{For a recent analysis, see N. Hata and P. Langacker,
 Phys. Rev. D50 (1994) 632.}
most likely,
\eqn\solar{
m_\nm^2-m_\ne^2\sim6\times10^{-6}\ eV^2,\ \ \
\sin^22\theta_{12}\sim7\times10^{-3}.}
In any case, the lepton parameters (certainly the charged lepton masses
and very likely also mixing angles and neutrino masses)
are small and hierarchical.

The smallness and hierarchy are the most puzzling features of
fermion parameters. They suggest that there should exist a more
fundamental theory where the hierarchy is generated in a natural
way. Once such a framework is found, we should investigate its
predictions and find ways in which it can be experimentally tested.
Such a task, based on the wealth of information on the
quark sector, was recently taken in refs.
\ref\lnsa{M. Leurer, Y. Nir and N. Seiberg, Nucl. Phys. B398 (1993) 319.}
\ref\qsa{Y. Nir and N. Seiberg, Phys. Lett. B309 (1993) 337.}
\ref\lnsb{M. Leurer, Y. Nir and N. Seiberg,
Nucl. Phys. B420 (1994) 468.}.
The smallness and hierarchy of the quark masses and mixing angles can be
naturally generated in theories which, at low energy, are described
effectively by an Abelian horizontal symmetry that is explicitly broken
by a small parameter. The possible evidence for such a symmetry
divides into three types:
\item{(i)} Numerology: the symmetry leads to order of magnitude
relations among the various parameters,
{\it e.g.} $|V_{ub}|\sim|V_{us}V_{cb}|$.
\item{(ii)} Direct evidence: the full high energy theory is likely
to require the existence of new particles, {\it e.g.} scalars with
flavor changing couplings and quarks in vector representations.
\item{(iii)} Indirect evidence: the symmetry has implications
for other sectors of the theory, {\it e.g.} the squark spectrum
\qsa, the Higgs potential \lnsb, leptoquark couplings
\ref\BaLe{E. Baver and M. Leurer, Phys. Rev. D51 (1995) 260.}
and baryon number violation
\ref\KaMa{H. Murayama and D.B. Kaplan, Phys. Lett. B336 (1994) 221.}
\ref\BeNi{V. Ben-Hamo and Y. Nir, Phys. Lett. B339 (1994) 77.}.

The conclusion of ref. \lnsb\ was that it is the indirect evidence
that is most likely to reveal the existence of the horizontal
symmetry mechanism. In this work we extend the study of refs.
\lnsa\qsa\lnsb\BaLe\BeNi\
to the lepton sector. (For recent, related work, see refs.
\ref\IbRo{L. Ib\'a\~nez and G.G. Ross, Phys. Lett. B332 (1994) 100.}
\ref\DLLRS{H. Dreiner, G.K. Leontaris, S. Lola, G.G. Ross and C. Scheich,
 CERN/TH.7412/94, hep-ph/9409369.}.)
Even though there is much less experimental information
in the lepton sector, we are able to find interesting predictions
that are independent of details of the lepton spectrum.

The paper is organized as follows: in section 2 we present the
theoretical low energy framework in which we work, namely the
lepton and scalar fields, the symmetry and the selection rules.
In section 3 we study features of the lepton parameters
that generally follow from the selection rules and are independent
of any specific phenomenological input. In section 4 we make some
assumptions about neutrino masses and mixing and present explicit
models where the resulting hierarchy in lepton parameters arises
naturally. In section 5 we investigate the possibility of exact
relations among lepton parameters. Section 6 studies the embedding
of the low energy effective model in a full high energy theory
and the constraints on the relevant energy scales. In section 7
we investigate whether lepton--slepton alignment could solve the
problem of unacceptably large lepton flavor changing processes without
slepton degeneracy. In section 8 we investigate the role of
horizontal symmetries in allowing for light leptoquarks.
In section 9, the possibility of an unbroken horizontal symmetry
is explored. Our conclusions are summarized in section 10.

\newsec{The Theoretical Framework}
We work in the framework of supersymmetric Abelian horizontal symmetry
that has been recently investigated in refs. \lnsa\qsa\lnsb. We assume
that the low energy spectrum consists of the fields of the minimal
supersymmetric Standard Model. The lepton and Higgs supermultiplets
carry the following $SU(2)_L\times U(1)_Y$ quantum numbers:
\eqn\gaugecharges{
L_i(2)_{-1/2},\ \ \bar\ell_i(1)_{+1},\ \ \phi_u(2)_{+1/2},\ \
\phi_d(2)_{-1/2}}
where $i=1,2,3$ is a generation index. The $L_i$ fields are
distinguished from the $\phi_d$ field by $R$ parity ($R_p$):
in $L_i$, the fermion components are $R_p$ even, while in $\phi_d$
the scalar component is $R_p$ even. Each of the supermultiplets
in \gaugecharges\ carries
a charge under an Abelian horizontal symmetry $\H$. For most of our
discussion, it makes no difference whether $\H$ is local or global,
continuous or discrete. (For a discrete $Z_N$, we usually assume that
$N$ is large enough that the Yukawa sector has an effective $U(1)$
symmetry.) $\H$ is explicitly broken by a small parameter $\l$
to which we attribute charge --1. Then, the following selection rules
apply:
\item{a.} Terms in the superpotential that carry charge $n\geq0$ under
$\H$ are suppressed by $\O(\l^n)$, while those with $n<0$ are forbidden
due to the holomorphy of the superpotential. (If $\H=Z_N$, the
suppression is by $\O(\l^{n(\mod N)})$.)
\item{b.} Terms in the K\"ahler potential that carry charge $n$ under
$\H$ are suppressed by $\O(\l^{|n|})$ (or $\O(\l^{{\rm min}[\pm n(\mod
N)]})$ for $\H=Z_N$).
\par
Explicitly, the lepton parameters arise from the Yukawa terms
\eqn\superpot{
Y_{ij}L_i\bl_j\phi_d+{Z_{ij}\over M_L}L_i L_j\phi_u\phi_u.}
$Y_{ij}$ is a generic complex $3\times3$ matrix that gives masses
to the charged leptons. $Z_{ij}$ is a symmetric complex $3\times3$
matrix that gives Majorana masses to the neutrinos. $M_L$ is
a high energy scale.

The first selection rule gives
\eqn\Yselect{H(L_i)+H(\bl_j)+H(\phi_d)=n\ \Longrightarrow
Y_{ij}=\cases{\O(\lambda^n)&$n\geq0$,\cr 0&$n<0$.}}
\eqn\Zselect{H(L_i)+H(L_j)+2H(\phi_u)=m\ \Longrightarrow
Z_{ij}=\cases{\O(\lambda^m)&$m\geq0$,\cr 0&$m<0$.}}
If the sum of charges is non-integer, the corresponding coupling
vanishes.

The second selection rule has implications for the
potential renormalization of the kinetic terms. The canonical terms
may be modified to
\eqn\kineticrenor{
\sum_{\ell,i,j}R^\ell_{ij}\ell_i^\dagger\gamma^\mu\partial_\mu\ell_j,}
where $\ell=L,\ \bl$ and $i,j=1,2,3$. Then
\eqn\Rselect{
H(\ell_i)-H(\ell_j)=k\ \Longrightarrow R^\ell_{ij}=\O(\l^{|k|}).}
The subtleties that arise when the kinetic terms are renormalized
were investigated in ref. \lnsb. It was proved that:
\item{1.} The coefficients of order one may change, but the
order of magnitude estimates in \Yselect\ and \Zselect\ remain valid.
\item{2.} The zeros in the Yukawa matrices may be lifted, but the
corresponding entries are still highly suppressed
and do not affect our results.

When discussing the lepton Yukawa couplings, we can set
two of the relevant $H$ charges to zero.
We can always use $U(1)_Y$ to set $H(\phi_u)=0$.
As \superpot\ is $U(1)_X$ invariant ($X(\phi_d)=-1$, $X(\bar\ell_i)=+1$,
$X(L_i)=X(\phi_u)=0$), we can set $H(\phi_d)=0$ without affecting
our analysis of the Yukawa sector. Note also that neutrino masses
depend on the unknown scale $M_L$. In models where all $H(L_i)$
are positive, we will use a modified scale,
$\tilde M={M_L\over\l^{2H(L_3)}}$, and modify the charge to $H(L_3)=0$.

Any horizontal symmetry that acts on the quark sector has to be
completely broken: an unbroken horizontal symmetry leads to either
degenerate quarks or vanishing mixing angles
\ref\Gatt{R. Gatto, G. Morchio and F. Strocchi, Phys. Lett. B83 (1979)
348; \hfill\break R. Gatto, G. Morchio, G. Sartori and F. Strocchi,
Nucl. Phys. B163 (1980) 221; \hfill\break
G. Segr\`e and H.A. Weldon, Phys. Lett. B86 (1979) 291; Ann. Phys. 124
(1980) 37.}\lnsa. Both possibilities
are experimentally excluded. On the other hand, as neither neutrino
masses nor lepton mixing angles have been unambiguously measured,
the possibility of an exact horizontal symmetry (acting non-trivially
in the lepton sector only) remains open. In most of this paper we
assume that $\H$ is completely broken, namely that all lepton
fields carry integer $H$-charges. We discuss the possibility
of an unbroken horizontal symmetry in section 9.

\newsec{General Results}
The special form of the neutrino Yukawa matrices \superpot,
which has no analogue in the charged fermion sectors,
has many interesting consequences. In particular, we find
that as far as ``numerology" is concerned, our framework
is much more predictive in the neutrino sector than in any
other sector.

There are nine independent physical parameters (ignoring CP-violation):
three mixing angles, three neutrino masses and three charged lepton
masses. However, as explained in the previous section,
there are only six relevant $H$-charges. Therefore, {\it our framework
predicts three order of magnitude relations among the physical
parameters independent of specific charge assignments.}\foot{
This situation should be compared to the quark sector,
where there is a single such relation, $|V_{ub}|\sim|V_{us}V_{cb}|$.}
These relations may involve the neutrino masses and the
mixing angles but not the charged lepton masses. The reason is that
the six mixing angles and neutrino masses depend on the three $H(L_i)$
only, while the three charged lepton masses depend also on the three
$H(\bl_i)$.

To find the three relations, we note that
the selection rules given in the previous section allow an
order of magnitude estimate for the various mass ratios and
mixing angles. With the choice $H(\l)=-1$, in most
of our models all lepton charges are positive, $H(L_i),H(\bl_i)\geq0$.
Then, ordering the lepton fields such that,
for $i<j$, $H(L_i)\geq H(L_j)$ and $H(\bl_i)\geq H(\bl_j)$, we get
\eqn\Omixing{\eqalign{
\sin\theta_{ij}\sim&\ \l^{H(L_i)-H(L_j)},\cr
{m_{\nu_i}\over m_{\nu_j}}\sim&\ \l^{2[H(L_i)-H(L_j)]},\cr
{m_{\ell_i}\over m_{\ell_j}}\sim&\ \l^{H(L_i)+H(\bl_i)-H(L_j)-H(\bl_j)}.
\cr}}
In some of our models, some lepton fields carry negative $H$ charges.
In these models, holomorphy plays an important role and the naive
estimates \Omixing\ may be violated. Consequently, the relations
given below do not hold in these models and other, model-specific
relations replace them.
The three relations can be easily found from Eq. \Omixing:
\eqn\Orelations{\eqalign{
{m_\ne\over m_{\nm}}\sim&\ \sin^2\theta_{12},\cr
{m_\nm\over m_{\nt}}\sim&\ \sin^2\theta_{23},\cr
{m_\ne\over m_{\nt}}\sim&\ \sin^2\theta_{13}.\cr}}
In other words, given, say, a single mass ratio in the neutrino sector
and a single (independent) mixing angle, our framework predicts the
order of magnitude of all other neutrino mass ratios and mixing angles.

Examining Eqs. \Omixing\ we further find\foot{
Similar relations should hold for quarks, namely
${m_{u_i}\over m_{u_j}},\ {m_{d_i}\over m_{d_j}}\lsim|V_{ij}|$.
It is encouraging that indeed all quark mass ratios fulfill
these inequalities.}
\eqn\mixlep{
\sin\theta_{ij}\gsim{m_{\ell_i}\over m_{\ell_j}}.}
As lepton masses are known, we can predict
\eqn\Omixnumbers{\eqalign{
\sin\theta_{12}\gsim&\ 0.002,\cr
\sin\theta_{23}\gsim&\ 0.03,\cr
\sin\theta_{13}\gsim&\ 0.0001.\cr}}
(To estimate these lower bounds, we take
${1\over2}{m_{\ell_i}\over m_{\ell_j}}$
to allow for the uncertainty in the order of magnitude estimates.)
Particularly encouraging is the rather large value of $\sin\theta_{23}$
which, if $m_\nt$ is in the appropriate range, would allow the detection
of $\nm-\nt$ oscillations in the CHORUS, NOMAD and E803 experiments.
Actually, this mixing angle should be rather close to the
experimental upper bound (for ``large $\Delta m_{23}^2$") \PDG:
$\sin^22\theta_{23}\lsim4\times10^{-3}$.

Eqs. \Omixing\ also imply (this is equivalent to the combination of
\Orelations\ and \mixlep)
\eqn\lepneu{
{m_{\nu_i}\over m_{\nu_j}}\gsim
\left({m_{\ell_i}\over m_{\ell_j}}\right)^2.}
We can predict then
\eqn\Oneunumbers{\eqalign{
{m_\ne\over m_\nm}\gsim&\ 1\times10^{-5},\cr
{m_\nm\over m_\nt}\gsim&\ 2\times10^{-3},\cr
{m_\ne\over m_\nt}\gsim&\ 4\times10^{-8}.\cr}}
(To estimate these lower bounds, we take
${1\over2}{m_{\ell_i}^2\over m_{\ell_j}^2}$
to allow for the uncertainty in the order of magnitude estimates.)
The prediction \lepneu\ has further interesting implications.
It coincides with the ``reasonable see-saw" assumption made in ref.
\ref\HaNi{H. Harari and Y. Nir, Phys. Lett. B188 (1987) 163;
Nucl. Phys. B292 (1987) 251.}.
In combination with cosmological constraints, it led to the
conclusion that all neutrinos are likely to be
lighter than $\O(100\ eV)$.

Finally, \Omixing\ implies that the hierarchy in the neutrino
sector is the same as in the charged lepton sector, namely
\eqn\noinvert{m_\ne\lsim m_\nm\lsim m_\nt.}
Here, $\ne,\nm,\nt$ denote the mass eigenstates with mixing
of $\O(1)$ with $e,\mu,\tau$, respectively.\foot{
Again, a similar result should hold for quarks and it does:
the CKM matrix is close to a unit matrix when we order
the up and down mass eigenstates in the same order of masses.}

\newsec{Explicit Examples}
For the purposes of this section, we will take the mass of
$\nt$ as given in \darkmatter, and the mass of $\nm$ and
the $\ne-\nm$ mixing angle as given in \solar. The charged
lepton masses are given in \chargedmasses. As we are interested
in explaining the orders of magnitude of the various parameters
and not in the exact numbers of order one, we will estimate the
various mass ratios and mixing angles as approximate powers of
a small parameter $\l$, where
\eqn\lnumber{
\l\sim0.2}
as follows from the value of the Cabibbo angle in the quark sector.
Explicitly:
\eqn\lcharged{
{m_e\over m_\mu}\sim\l^3,\ \ \ {m_\mu\over m_\tau}\sim\l^2,}
\eqn\lneutral{
{m_\nm\over m_\nt}\sim\l^{5\pm1},\ \ \ \sin\theta_{12}\sim\l^2.}
We will further assume that $\tan\beta\equiv{\vev{\phi_u}\over
\vev{\phi_d}}\sim1$ and, consequently, ${m_\tau\over\vev{\phi_d}}
\sim\l^3$. It is simple to modify our models to the case of
large $\tan\beta$.

In the simplest case, the horizontal symmetry and the charge of
the breaking parameter are
\eqn\HUone
{\H=U(1),\ \ \ H(\l)=-1.}
As explained above, the input data in \lcharged\ and \lneutral\
are enough to fix the $H$-charges of all the relevant fields
and, therefore, predict the order of magnitude of all remaining
lepton parameters. If $m_\nm/m_\nt\sim\l^5$, there is an
unbroken horizontal $Z_2$ symmetry, $\tau$-parity. We discuss
this case separately in section 9.

\subsec{$m_\nm/m_\nt\sim\l^4$}
The hierarchy in Eqs. \lcharged\ and \lneutral\ determines
a unique set of $H$ charge assignments:
\eqn\fourcharge{\matrix{
L_1&L_2&L_3&&\bar\ell_1&\bar\ell_2&\bar\ell_3&&\phi_u&\phi_d\cr
(4)&(2)&(0)&&(4)&(3)&(3)&&(0)&(0).\cr}}
The lepton mass matrices have then the following form:
\eqn\fourmasses{
M^\ell\sim\vev{\phi_d}
\pmatrix{\l^8&\l^7&\l^7\cr\l^6&\l^5&\l^5\cr\l^4&\l^3&\l^3\cr},
\ \ \ M^\nu\sim{\vev{\phi_u}^2\over\tM}
\pmatrix{\l^8&\l^6&\l^4\cr\l^6&\l^4&\l^2\cr\l^4&\l^2&1\cr}.}
We emphasize that, here and below, the sign ``$\sim$" implies that we
only give the order of magnitude of the various entries; there is an
unknown (complex) coefficient of $\O(1)$ in each entry that we do
not write explicitly. Eq. \fourmasses\
predicts
\eqn\fourpredict{
\sin\theta_{23}\sim\l^2,\ \ \ \sin\theta_{13}\sim\l^4,
\ \ \ m_\ne/m_\nm\sim\l^4,}
consistent with \Orelations. Note, in particular, that
\eqn\fourchorus{
m_\nt^2-m_\nm^2\sim100\ eV^2,\ \ \ \sin^22\theta_{23}\sim0.006,}
which is of the order of present bounds.

\subsec{$m_\nm/m_\nt\sim\l^6$}
This is an example of a model where
\eqn\violate{
{m_\nm\over m_\nt}<\left({m_\mu\over m_\tau}\right)^2.}
This violates the naive prediction \lepneu. As explained in section 3,
\violate\ can only be accommodated in models where
some of the horizontal charges are negative and holomorphy
then gives zero entries in the mass matrices. Specifically,
for the case under study, $M_{32}^\ell=0$ is required, so that
at least one of the $H_i$-charges has to fulfill
$H_i(L_3)+H_i(\bar\ell_2)<0$.

As a specific example we take
\eqn\Hsix{
H=U(1)_{H_1}\times U(1)_{H_2};\ \ \ \l_1(-1,0)\sim\l,\ \
\l_2(0,-1)\sim\l^2.}
Take the following set of $(H_1,H_2)$ charges:
\eqn\sixncharge{\matrix{
L_1&L_2&L_3&&\bar\ell_1&\bar\ell_2&\bar\ell_3&&\phi_u&\phi_d\cr
(1,2)&(3,0)&(0,0)&&(-1,2)&(-2,2)&(1,1)&&(0,0)&(0,0).\cr}}
The lepton mass matrices have the following form:
\eqn\sixnmasses{
M^\ell\sim\vev{\phi_d}\pmatrix{\l_2^4&0&\l_1^2\l_2^3\cr
\l_1^2\l_2^2&\l_1\l_2^2&\l_1^4\l_2\cr0&0&\l_1\l_2\cr},
\ \ \ M^\nu\sim{\vev{\phi_u}^2\over\tilde M}
\pmatrix{\l_1^2\l_2^4&\l_1^4\l_2^2&\l_1\l_2^2\cr
\l_1^4\l_2^2&\l_1^6&\l_1^3\cr\l_1\l_2^2&\l_1^3&1\cr}.}
It predicts:
\eqn\sixpredict{
\sin\theta_{23}\sim\l^3,\ \ \ \sin\theta_{13}\sim\l^5,
\ \ \ m_\ne/m_\nm\sim\l^4.}
In this example, the relations \Orelations\ and \noinvert\
are maintained, but \mixlep\ and \lepneu\ are circumvented
due to the holomorphy of the superpotential.

\newsec{Exact Relations}
We investigate the possibility that the horizontal symmetry dictates
not just order of magnitude relations but also exact (typically
to order $\l$ or $\l^2$) relations. That can be the case if there
are enough zeros in the mass matrices. These zeros arise from the
holomorphy of the superpotential, when a certain entry breaks $\H$
by a {\it negative} charge.

Exact predictions arise trivially in the case that
$m_\nm/m_\nt\sim\l^5$ as will be discussed in section 9. More generally,
if any ratio between neutrino masses is an odd power of the
small breaking parameter, then the mixing angle between the
two neutrinos will vanish. This is a result of a symmetry and
therefore exact to all orders.

We now investigate the possibility of exact relations
in the case $m_\nm/m_\nt\sim\l^4$, described in section 4.1.
We used the techniques for diagonalizing the mass matrices
described in refs.
\ref\HaRa{L.J. Hall and A. Rasin, Phys. Lett. B315 (1993) 164.}\lnsb.
Let us recall that the mixing angles are given by ($s_{ij}\equiv
\sin\theta_{ij}$)
\eqn\mixingangles{\eqalign{
s_{12}=&\ |s_{12}^\ell-s_{12}^\nu|,\cr
s_{23}=&\ |s_{23}^\ell-s_{23}^\nu|,\cr
s_{13}=&\ |s_{13}^\ell-s_{13}^\nu-s_{12}^\nu(s_{23}^\ell-s_{23}^\nu)|,
\cr}}
where $s_{ij}^\nu$ ($s_{ij}^\ell$) are the elements of $V_L^\nu$
($V_L^\ell$) where
\eqn\diagonalize{\eqalign{
V_L^\nu M^\nu(V_L^\nu)^T=&\ {\rm diag}(m_\ne,m_\nm,m_\nt),\cr
V_L^\ell M^\ell(V_R^\ell)^\dagger=&\ {\rm diag}(m_e,m_\mu,m_\tau).\cr}}

For the neutrinos, we define
\eqn\yijn{\eqalign{y_{ij}^\nu\equiv&\ M_{ij}^\nu/M_{33}^\nu,\cr
\tilde y_{22}^\nu=&\ y_{22}^\nu-y_{23}^\nu y_{32}^\nu.\cr}}
(Note: $y_{33}^\nu=1$; $|\tilde y^\nu_{22}|=m_\nm/m_\nt$.)
The $s_{ij}^\nu$ mixing angles are
(we write only the potentially leading terms for $M^\nu$ of \fourmasses)
\eqn\sijnu{
s_{12}^\nu={y_{12}^\nu\over\tilde y_{22}^\nu}-{y_{13}^\nu
 y_{32}^\nu\over\tilde y_{22}^\nu},\ \ \
s_{13}^\nu= y_{13}^\nu,\ \ \
s_{23}^\nu= y_{23}^\nu.}

For the charged leptons, we define
\eqn\yijl{\eqalign{
y_{i1}^\ell=&{M^\ell_{i1}\over\sqrt{|M_{32}^\ell|^2+|M_{33}^\ell|^2}},\cr
y_{i2}^\ell=&{M^\ell_{i2}M^\ell_{33}-M^\ell_{i3}M^\ell_{32}
\over|M_{32}^\ell|^2+|M_{33}^\ell|^2},\cr
y_{i3}^\ell=&{M^\ell_{i3}M^\ell_{33}-M^\ell_{i2}M^\ell_{32}
\over|M_{32}^\ell|^2+|M_{33}^\ell|^2}.\cr}}
The $s_{ij}^\ell$ mixing angles are
(we write only the potentially leading terms for $M^\ell$ of \fourmasses)
\eqn\sijel{
s_{12}^\ell={y_{12}^\ell\over y_{22}^\ell},\ \ \
s_{13}^\ell=y_{13}^\ell,\ \ \
s_{23}^\ell=y_{23}^\ell.}

We find two possible exact relations.
First, take mass matrices of the form
\eqn\amasses{
M^\ell\sim\vev{\phi_d}
\pmatrix{\l^8&0&\l^7\cr\l^6&\l^5&\l^5\cr\l^4&0&\l^3\cr},
\ \ \ M^\nu\sim{\vev{\phi_u}^2\over M_L}
\pmatrix{0&\l^6&0\cr\l^6&\l^4&\l^2\cr0&\l^2&1\cr}.}
With $M^\nu_{13}=0$ we have $s_{13}^\nu=0$, while with
$M^\ell_{12}=M^\ell_{32}=0$ we have $s_{12}^\ell=0$.
Then
\eqn\aspacial{{m_\ne\over m_\nm}=s_{12}^2.}
An example of charges that lead
to \amasses\ is a symmetry of the type \Hsix\ with
\eqn\acharge{\matrix{
L_1&L_2&L_3&&\bar\ell_1&\bar\ell_2&\bar\ell_3&&\phi_u&\phi_d\cr
(6,-1)&(0,1)&(0,0)&&(2,1)&(5,-1)&(1,1)&&(0,0)&(0,0).\cr}}

Second, take mass matrices of the form
\eqn\cmasses{
M^\ell\sim\vev{\phi_d}
\pmatrix{\l^8&\l^7&\l^7\cr\l^6&\l^5&0\cr\l^4&0&\l^3\cr},
\ \ \ M^\nu\sim{\vev{\phi_u}^2\over M_L}
\pmatrix{\l^{10}&\l^8&\l^6\cr\l^8&0&\l^4\cr\l^6&\l^4&\l^2\cr}.}
With $M^\nu_{22}=0$ we have $m_\nm/m_\nt=(s_{23}^\nu)^2$, while with
$M^\ell_{23}=M^\ell_{32}=0$ we have $s_{23}^\ell=0$. Then
\eqn\cspacial{{m_\nm\over m_\nt}=s_{23}^2.}
An example of charges that lead
to \cmasses\ is a symmetry of the type \Hsix\ with
\eqn\ccharge{\matrix{
L_1&L_2&L_3&&\bar\ell_1&\bar\ell_2&\bar\ell_3&&\phi_u&\phi_d\cr
(1,2)&(-1,2)&(1,0)&&(1,1)&(4,-1)&(0,1)&&(0,0)&(0,0).\cr}}

To summarize, the generic order of magnitude relations between
neutrino mass ratios and mixing angles \Orelations\ may be
promoted to exact relations (for $m_\ne/m_\nm$ or $m_\nm/m_\nt$)
if holomorphy induces certain zero entries in the lepton
mass matrices.

\newsec{A Full High Energy Theory}
So far we have considered a low energy effective theory with
an explicitly broken horizontal symmetry $\H$ that gives certain
selection rules. The most natural full high energy theory where
our model can be embedded is that of Froggatt and Nielsen
\ref\FrNi{C.D. Froggatt and H.B. Nielsen, Nucl. Phys. B147 (1979) 277.}.
$\H$ is an exact symmetry that is spontaneously broken by a
vev $\vev{S}$ of a scalar field that is a singlet under the
SM gauge group.
The information about the breaking is communicated to the
observed leptons by heavy leptons in vector representations.
The mass scale for these lepton supermultiplets is denoted by
$M$, and the small parameter $\l=\vev{S}/M$.
The relevant $SU(2)_L\times U(1)_Y\times U(1)_H$
representations of the heavy leptons are
\eqn\mirror{\eqalign{
F(2,-1/2,H)\ \ {\rm and}&\ \ \bar F(2,+1/2,-H),\cr
E(1,-1,H)\ \ {\rm and}&\ \ \bar E(1,+1,-H),\cr
N(1,0,H)\ \ {\rm and}&\ \ \bar N(1,0,-H).\cr}}

An important question is whether any of these new particles can be
observed in experiment, thus providing direct evidence for the
horizontal symmetry framework. The answer depends on the scales
$\vev{S}$ and $M$: the low energy parameters determine only the
ratio between them. If they are low enough, say $M\lsim500\ TeV$,
then the particles may, in principle, be accessible in
future experiments. Lower bounds on these scales come
from flavor changing neutral currents (FCNC) and from requiring that no
Landau poles appear in the running of the gauge couplings below the
Planck scale \lnsa\lnsb.

The scalar $S$ mediates various lepton flavor violating processes,
{\it e.g.} $\mu\ra eee$ and $K\ra\pi\mu e$. We assume that the mass
of $S$ is of the order of the SUSY breaking scale (see the analysis
of the Higgs potential in ref. \lnsb), and that its coupling to
a fermion pair $f_i\bar f_j$ is of order ${(m_i+m_j)\sin\theta_{ij}
\over\vev{S}}$. Surveying all relevant FCNC processes (for a recent
review, see
\ref\KLV{T.S. Kosmas, G.K. Leontaris and J.D. Vergados,
 Prog. Part. Nucl. Phys. 33 (1994) 397.}),
we find that all the bounds on $\vev{S}$
are below the electroweak scale. Similarly, the new leptons
contribute to lepton flavor violating processes through loop
diagrams, but no significant bounds arise.

The $N,\bar N$ supermultiplets do not affect the running of the
gauge couplings. The number of $F$-doublets $N_F$ (=$N_{\bar F}$)
and the number of $E$-singlets $N_E$ (=$N_{\bar E}$) affect the
running of the $SU(2)_L$ and $U(1)_Y$ coupling constants and
may lead to Landau poles. The relevant counting is \lnsb\
\eqn\ntwonone{\eqalign{
N_2=&\ 3N_P+N_F,\cr
N_1=&\ N_P+8N_U+2N_D+3N_F+6N_E,\cr}}
where $N_P$ is the number of quark doublets and
$N_U(N_D)$ is the number of quark singlets
of charge +2/3(--1/3). If, for example, we are interested
in $M\leq400\ TeV$, then $N_2\leq5$ and $N_1\leq18$.

The minimal number of heavy leptons required can be easily deduced from
the suppression of the light masses. For the model defined by
\fourcharge, Eq. \fourmasses\ gives
$\det M^\ell\sim\vev{\phi_d}^3\l^{16}$, which requires $N_F+N_E\geq16$,
(if $m_\tau/\vev{\phi_d}\sim1$, $\det M^\ell\sim\vev{\phi_d}^3\l^7$
and $N_F+N_E\geq7$),
and $\det M^\nu\sim{\vev{\phi_u}^6\over M^3}\l^{12}$, which requires
$N_F+N_N\geq12$ and $N_{\hat N}=3$ where $\hat N$ are $\H$-singlets
($\hat N(1,0,0)$).
We will assume that all heavy neutrinos are $SU(2)_L$ singlets,
in which case they are irrelevant to the investigation of Landau poles.
There is no way to have $N_F+N_E\geq7$,
$N_F\leq N_2\leq5$ and $3N_F+6N_E\leq N_1\leq18$.
Therefore, if the smallness of lepton masses is a result of
a horizontal $U(1)_H$ symmetry, the breaking of the symmetry
is at an energy too high for direct observation.

The lowest possible scale in our framework is allowed when $H=U(1)_{H_1}
\times U(1)_{H_2}$, $\l_1\sim\l^2$ and $\l_2\sim\l^3$.
Here we could imagine $m_\mu/m_\tau\sim\l_1$, $m_e/m_\tau\sim\l_1\l_2$
and $m_\tau/\vev{\phi_d}\sim1$.
In such a case, only 3 heavy charged leptons are needed, so we can have
$N_F=2$ and $N_E=1$. As shown in ref. \lnsb, this -- together
with a specific choice of quark representations --
allows $M\geq900\ TeV$.

To summarize, a natural embedding of our low energy effective theory
in a full high energy framework  requires the existence of scalars
with flavor changing couplings and new leptons in vector representations.
Constraints on the masses of these new particles from FCNC are very mild.
However, if the gauge group is that of the Standard Model up to
the Planck scale, then Landau poles constraints make it very unlikely
that the new particles are light enough to be directly observed in
future experiments.

We would like to emphasize the following point
\ref\RaSi{A. Rasin and J.P. Silva, Phys. Rev. D49 (1994) 20.}:
{\it Our various
predictions for the light neutrino parameters, e.g. the predictions
in section 3, are independent of the structure of the right-handed
neutrino mass matrix.} This is in contrast to most models of
see-saw mass matrices where one has to assume a specific hierarchy
in the mass matrix for the heavy neutrinos in order to have any
prediction for the light spectrum. The point can be proved as follows.
Denote by $M_R$ the Majorana mass matrix for the heavy $N$ fields,
and by $M_D$ the Dirac mass matrix. Then
\eqn\seesaw{[M^\nu]_{ij}\sim [M_D]_{ik}[(M_R)^{-1}]_{kl}[M_D^T]_{lj}.}
Assume that $N_k$ carries horizontal charge $H_k$. Then
$[M_D]_{ik}\propto\l^{H_i+H_k}$,
$[M_D]_{jl}\propto\l^{H_j+H_l}$, and
$[M_R]_{kl}\propto\l^{H_k+H_l}$.
Our selection rules further imply that $[(M_R)^{-1}]_{kl}\sim
([M_R]_{kl})^{-1}$. Then the product
$[M_D]_{ik}[(M_R)^{-1}]_{kl}[M_D^T]_{lj}$ is independent of both $H_k$
and $H_l$. {\it Our low energy predictions are valid for arbitrary
structure of $M_R$} (as long as it has no zero eigenvalues).

\newsec{Lepton Slepton Alignment}
For generic slepton masses, penguin diagrams with sleptons and
neutralinos give unacceptably large contributions to radiative
charged lepton decays
\ref\GaMa{F. Gabbiani and A. Masiero, Nucl. Phys. B322 (1989) 235;\hfill
\break D. Choudhury, F. Eberlein, A. K\"onig, J. Louis and
 S. Pokorski, Phys. Lett. B342 (1995) 180.}.
The standard solution to this problem
is to assume that sleptons are degenerate to a very good approximation.
This is not motivated in generic supergravity models or string
theory, though it may hold under special conditions
\ref\KaLo{V. Kaplunovski and J. Louis, Phys. Lett. B306 (1993) 451;
\hfill\break R. Barbieri, J. Louis and M. Moretti, Phys. Lett.
B312 (1993) 451; \hfill\break J. Louis and Y. Nir, Munich preprint
LMU-TPW 94-17, hep-ph/9411429.}.
Both slepton degeneracy and proportionality of trilinear
Higgs--slepton couplings to Yukawa couplings can be natural if
Supersymmetry breaking is communicated to the light particles
by gauge interactions
\ref\DKS{M. Dine, A. Kagan and S. Samuel, Phys. Lett. B243 (1990) 250;
\hfill\break M. Dine and A. Nelson, Phys. Rev. D48 (1993) 1277.},
or in models with a non-Abelian horizontal symmetry
\ref\DKL{M. Dine, A. Kagan and R. Leigh, Phys. Rev. D48 (1993) 4269.}.

Recently, an alternative mechanism has been suggested to solve
the FCNC problem in the quark sector \qsa\lnsb: the approximate
alignment of quark mass matrices with squark mass-squared matrices.
In this section we investigate whether such a mechanism can work
for the lepton sector as well. The idea is that a horizontal symmetry,
of the type discussed in this work, forces both $M^f$ and
$\tilde M^{f2}$ ($f=\ell,\nu$)
to be approximately diagonal in the basis where
the horizontal  charges are well defined. Consequently, the
mixing matrix for lepton--slepton--photino couplings is close to
a unit matrix and FCNCs are suppressed, regardless of whether
sleptons are degenerate or not.

To explicitly investigate this mechanism, we define diagonalizing
matrices $V_M^f$ and $\tilde V_M^f$:
\eqn\diagV{\eqalign{
V_L^\ell M^\ell V_R^{\ell\dagger}=&\ {\rm diag}\{m_e,m_\mu,m_\tau\},\cr
V_L^\nu M^\nu V_L^{\nu T}=&\ {\rm diag}\{m_\ne,m_\nm,m_\nt\},\cr}}
\eqn\diagtV{\eqalign{\tV_L^f\tM^{f2}_{LL}\tV_L^{f\dagger}=&\ {\rm diag}
\{m^2_{\tilde f_{L1}},m^2_{\tilde f_{L2}},m^2_{\tilde f_{L3}}\},\cr
\tV_R^\ell\tM^{\ell2}_{RR}\tV_R^{\ell\dagger}=&\ {\rm diag}
\{m^2_{\tilde e_{R}},m^2_{\tilde\mu_{R}},m^2_{\tilde\tau_{R}}\}.\cr}}
We assume here that the average slepton mass, $m_{\tilde\ell}$ and
$m_{\tilde\nu}$, is somewhat higher than the electroweak breaking
scale $m_Z$, so that $\tM_{LL}^{\ell2}\approx\tM_{LL}^{\nu2}$.
The neutralino mixing matrices are then
\eqn\neutralino{K_L^f=V_L^f\tV_L^{f\dagger},\ \ \
K_R^\ell=V_R^\ell\tV_R^{\ell\dagger}.}
Supersymmetric contributions to FCNC processes will be proportional to
\eqn\defdeltaij{(\delta_{MM}^f)_{ij}\sim\max_\alpha(K_M^f)_{i\alpha}
(K_M^f)^*_{j\alpha}.}
Furthermore, the off-diagonal blocks are suppressed by
$\O(m_f/m_{\tilde f})$ compared to the diagonal blocks.
Their contributions to FCNC are then proportional to
\eqn\deltaLR{(\delta^\ell_{LR})_{ij}=(V_L^\ell\tM^{\ell2}_{LR}
V_R^{\ell\dagger})_{ij}/m_{\tilde\ell}^2.}

We have checked the bounds on $(\delta^f_{MN})_{ij}$ from all
relevant lepton flavor changing processes. The strongest bounds come from
radiative lepton decays. We find \GaMa\
(using $m_{\tilde\gamma}^2/m_{\tilde\ell}^2=0.5$):
\eqn\meg{
BR(\mu\rightarrow e\gamma)\leq4.9\times10^{-11}\Longrightarrow\cases{
(\delta_{MM}^\ell)_{12}\leq&$1\times10^{-1}\ \left({m_{\tilde\ell}
\over1\ TeV}\right)^2$,\cr
(\delta_{LR}^\ell)_{12}\leq&$2\times10^{-5}\ \left({m_{\tilde\ell}
\over1\ TeV}\right)$.\cr}}
\eqn\teg{
BR(\tau\rightarrow e\gamma)\leq2.0\times10^{-4}\Longrightarrow\cases{
(\delta_{MM}^\ell)_{13}\leq&$4\times10^{3}\ \left({m_{\tilde\ell}
\over1\ TeV}\right)^2$,\cr
(\delta_{LR}^\ell)_{13}\leq&$2\ \left({m_{\tilde\ell}
\over1\ TeV}\right)$.\cr}}
\eqn\tmg{
BR(\tau\rightarrow\mu\gamma)\leq4.2\times10^{-6}\Longrightarrow\cases{
(\delta_{MM}^\ell)_{23}\leq&$7\times10^{2}\ \left({m_{\tilde\ell}
\over1\ TeV}\right)^2$,\cr
(\delta_{LR}^\ell)_{23}\leq&$2\times10^{-1}\ \left({m_{\tilde\ell}
\over1\ TeV}\right)$.\cr}}

For the $\delta_{LR}^\ell$ matrix, the horizontal symmetry implies
\eqn\MellLR{(\tilde M^{\ell2}_{LR})_{ij}\sim m_{\tilde\ell}M^\ell_{ij},}
which gives, with the choice \fourcharge,
\eqn\delLR{\eqalign{
(\delta_{LR}^\ell)_{12}\sim&\ {m_\mu\sin\theta_{12}\over m_{\tilde\ell}}
\sim4\times10^{-6}\left({1\ TeV\over m_{\tilde\ell}}\right),\cr
(\delta_{LR}^\ell)_{13}\sim&\ {m_\tau\sin\theta_{13}\over m_{\tilde\ell}}
\sim3\times10^{-6}\left({1\ TeV\over m_{\tilde\ell}}\right),\cr
(\delta_{LR}^\ell)_{23}\sim&\ {m_\tau\sin\theta_{23}\over m_{\tilde\ell}}
\sim7\times10^{-5}\left({1\ TeV\over m_{\tilde\ell}}\right).\cr}}
We see that the constraints on $(\delta^\ell_{LR})_{ij}$ are
satisfied. (For $\mu\rightarrow e\gamma$,
\meg\ and \delLR\ imply $m_{\tilde\ell}\geq200\ GeV$.)

For the $\delta_{MM}^\ell$ matrices, the $(\delta^\ell_{MM})_{i3}$
elements are not really constrained. On the other hand,
\eqn\megl{
(\delta_{MM}^\ell)_{12}\lsim\ \l^2}
is required. (For $m_{\tilde\ell}\gsim1\ TeV$ the bound is relaxed
to $\l$.) As we require $\sin\theta_{12}\sim\l^2$ and $m_e/m_\mu
\sim\l^3$, the naive estimates are
\eqn\Hmegl{
(\delta_{LL}^\ell)_{12}\lsim\ \l^2,\ \ \
(\delta_{RR}^\ell)_{12}\lsim\ \l.}
The only potential problem is then with $(\delta_{RR}^\ell)_{12}$,
and even this is automatically solved if $m_{\tilde\ell}$ is
somewhat on the heavy side. Below we present an example of
a horizontal symmetry that aligns leptons
and sleptons of the first two generations so that
$(K_R^\ell)_{12}$ is satisfactorily suppressed even for light sleptons.

Take $\H=U(1)_{H_1}\times U(1)_{H_2}$,
$\l_1\sim\l$ and $\l_2\sim\l^2$, and charge assignments
\eqn\lsacharge{\matrix{
L_1&L_2&L_3&&\bar\ell_1&\bar\ell_2&\bar\ell_3&&\phi_u&\phi_d\cr
(0,2)&(2,0)&(0,0)&&(6,-1)&(1,1)&(1,1)&&(0,0)&(0,0).\cr}}
It leads to the following mass matrices for charged leptons
and sleptons:
\eqn\lsamasses{
M^\ell\sim\vev{\phi_d}\pmatrix{\l_1^6\l_2&\l_1\l_2^3&\l_1\l_2^3\cr
0&\l_1^3\l_2&\l_1^3\l_2\cr 0&\l_1\l_2&\l_1\l_2\cr},}
\eqn\lsasmass{
\tilde M^{\ell2}_{LL}\sim m_{\tilde\ell}^2\pmatrix{
1&\l_1^2\l_2^2&\l_2^2\cr \l_1^2\l_2^2&1&\l_1^2\cr \l_2^2&\l_1^2&1\cr},\ \
\tilde M^{\ell2}_{RR}\sim m_{\tilde\ell}^2\pmatrix{
1&\l_1^5\l_2^2&\l_1^5\l_2^2\cr \l_1^5\l_2^2&1&1\cr \l_2^5\l_2^2&1&1\cr}.}
In this model
\eqn\lsamixing{\eqalign{
(V_L^\ell)_{12}\sim\l^2,\ \ (\tilde V^\ell_L)_{12}\sim\l^6\ \ &
\Longrightarrow (\delta_{LL}^\ell)_{12}\sim\l^2,\cr
(V_R^\ell)_{12}\sim\l^5,\ \ (\tilde V^\ell_R)_{12}\sim\l^9\ \ &
\Longrightarrow (\delta_{RR}^\ell)_{12}\sim\l^5,\cr}}
and all constraints are satisfied.

To summarize, the horizontal symmetry aligns lepton mass matrices
with slepton mass-squared matrices.
If slepton masses are of $\O(1\ TeV)$, then SUSY penguin diagrams
contributions to radiative lepton decays are below the experimental
bounds and no slepton degeneracy needs to be assumed.
If sleptons are lighter than $\O(1\ TeV)$, then either
${m^2_{\tilde e_R}-m^2_{\tilde \mu_R}\over m^2_{\tilde\ell_R}}\lsim0.2$,
or the combination of horizontal symmetry and holomorphy produces
an alignment that is more precise than the naive estimate.

\newsec{Naturally Light Leptoquarks}
Light ($M_{LQ}=\O(TeV)$) leptoquarks may be discovered in the DESY $ep$
collider HERA and in Fermilab. However, light leptoquarks pose severe
phenomenological problems unless their couplings are chiral and diagonal
(see {\it e.g.} ref.
\ref\Leurer{M. Leurer, Phys. Rev. Lett. 71 (1993) 1324;
 Phys. Rev. D49 (1994) 333.}).
Horizontal symmetries are necessary to guarantee these features in
a natural way \BaLe. In the analysis of ref. \BaLe, it was assumed
that neutrinos are massless. In
this section we extend their analysis to the case of massive neutrinos.
Also, while ref. \BaLe\ presented a single example, we show that actually
a rather large class of models allows light leptoquarks.

Following ref. \BaLe, we focus on the case motivated by $E_6$ (or
superstring) models, where the leptoquark supermultiplets are in
the following $SU(3)_C\times SU(2)_L\times U(1)_Y$ representations:
\eqn\LQgauge{S(\bar 3,1)_{+1/3},\ \ \ \SP(3,1)_{-1/3}.}
In the superpotential, the following renormalizable couplings appear:
\eqn\LQSP{(G_L)_{ij}L_iQ_jS+(G_R)_{ij}\bar\ell_i\bar u_j S^\prime}
(where $Q(3,2)_{+1/6}$ and $\bar u(\bar3,1)_{-2/3}$ are quark
supermultiplets). The phenomenologically relevant quantities are
the coupling matrices in the mass basis,
\eqn\LQGmass{G^{\nu d}_L=V^{\nu\dagger}G_L V^d_L,\ \ \
G^{\ell u}_L=V^{\ell\dagger}_L G_L V^u_L,\ \ \
G^{\ell u}_R=V^{\ell\dagger}_R G_R V^u_R.}
(The matrices $V^{d,u}_{L,R}$ are defined analogously to the
lepton diagonalizing matrices, see Eq. \diagV.)

We are interested in finding models where the $S$ leptoquark
is allowed to have a light mass while its diagonal coupling to first
generation fermions is not suppressed, namely
\eqn\desiredLQ{M_S=\O(TeV),\ \ \ (G^{\nu d}_L)_{11}=\O(1),
\ \ \ (G_L^{\ell u})_{11}=\O(1).}
This task is difficult because off-diagonal couplings have to
be highly suppressed \BaLe\Leurer
\ref\DBC{S. Davidson, D. Bailey and B.A. Campbell,
 Z. Phys. C61 (1994) 613.}. In particular,
$\mu-e$ conversion in nuclei requires
\eqn\Glubound{
(G^{\ell u}_L)_{21}\leq10^{-4}\left({M_S\over1\ TeV}\right),}
while $K-\bar K$ mixing requires
\eqn\Gndbound{(G^{\nu d}_L)_{12}\leq3\times10^{-2}
\left({M_S\over1\ TeV}\right).}
Finally, we need to allow a mixing term in the superpotential
for the $S$ and $\SP$ field, $\mu_{S\SP}S\SP$, to give masses to
the fermionic components, but then the $G_R$ matrix has to be
highly suppressed to avoid inconsistency with the measured leptonic
decays of the pion:
\eqn\GRbound{\mu_{S\SP}\gsim m_Z,\ \ \ (G_R^{\ell u})_{ij}\ll1.}

A horizontal symmetry is essential in naturally fulfilling these
requirements in three important ways:
\item{a.} In the {\it interaction} basis, $\H$ can give
\eqn\HonGint{G_L\sim\pmatrix{1&0&0\cr 0&0&0\cr 0&0&0\cr},\ \ \
G_R\sim\pmatrix{0&0&0\cr 0&0&0\cr 0&0&0\cr},}
where by zero entry we mean that it is highly suppressed ($\sim\l^n$
with $n\gg1$).
\item{b.} In the {\it mass} basis, the off diagonal couplings
are still small because the horizontal symmetry forces the
lepton mass matrices and quark mass matrices to be
simultaneously approximately diagonal.
\item{c.} The horizontal symmetry together with holomorphy
can make specific couplings particularly small (similar to
the quark squark alignment where gluino couplings in the
down sector are highly suppressed).

We now explain these points in more detail.

$a.$ In order to make $(G_L)_{11}$ the only unsuppressed
leptoquark coupling, we can choose
\eqn\LQint{H(S)+H(Q_1)+H(L_1)=0;\ \ \ H(\SP)=-H(S).}
As $H(Q_i)+H(L_j)<H(Q_1)+H(L_1)$ for $i$ and/or $j\neq1$,
$(G_L)_{ij}$ would either vanish (due to holomorphy) or
be highly suppressed if $\H$ is discrete. Also
$(G_R)_{ij}\sim\l^{H(Q_1)+H(L_1)+H(\bar\ell_i)+H(\bar u_j)}$
would be highly suppressed. Effectively, we get \HonGint.

Three comments are in order:
\par (i) The term $\mu_{S\SP}S\SP$ in
the superpotential, which is necessary to give
the fermionic components of $S$ and $\SP$ masses, is allowed.
\par (ii) The $(G_R)_{ij}$ couplings can be further suppressed
by choosing horizontal charges $\tilde H=H+\alpha X$
where $X(\phi_d)=-X(\bar d_i)=-X(\bar\ell_i)$ and all
other fields carry $X=0$. $U(1)_X$ is an accidental symmetry
of the Yukawa matrices. Consequently, $H$ and $\tilde H$
are isomorphic as far as fermion masses and mixings are concerned.
\par (iii) Another way to suppress the $G_R$ matrix is by
taking $H(\SP)>-H(S)$. In this way also $\mu_{S\SP}$ is
suppressed, but its scale is anyway unknown.

$b.$ The horizontal symmetry guarantees that the
various elements in the diagonalizing matrices are
not larger than the corresponding elements in the
mixing matrices, {\it e.g.} $(V^u_L)_{12},(V^d_L)_{12}\lsim
V_{us}$. ($V$ here is the CKM mixing matrix for quarks.
Below we denote the charged current mixing matrix for leptons
by $U$.) Consequently,
\eqn\HonGmass{\eqalign{
(G_L^{\ell u})_{ij}\sim&\ (V_L^\ell)_{1i}(V_L^u)_{1j}
\lsim U_{1i}V_{1j},\cr
(G_L^{\nu d})_{ij}\sim&\ (V_L^\nu)_{1i}(V_L^d)_{1j}
\lsim U_{1i}V_{1j}.\cr}}
Thus, for example, as $V_{12}\sim\l$ and we assume $U_{12}\sim\l^2$,
the bound on $M_S$ from \Glubound\ relaxes by a factor of $\l^2$
to $M_S\gsim400\ TeV$, while
the bound \Gndbound\ relaxes by a factor of $\l$
to $M_S\gsim8\ TeV$.

$c.$ The strongest bounds come from the charged lepton and
down sectors. Thus, the lower bounds on $M_S$ would be weakest
in models where
\eqn\MSsmall{(V^d_L)_{12}\ll V_{12}\sim\l,\ \ \
(V^\ell_L)_{12}\ll U_{12}\sim\l^2.}
These are precisely the models of quark-squark alignment
(see ref. \lnsb) and lepton-slepton alignment (see the
previous section). By using quark squark alignment, the
bound from $K-\bar K$ mixing is avoided, and only a weaker
bound from $D-\bar D$ mixing holds:
\eqn\LQQSA{M_S\geq\O(5\ TeV).}
By using lepton slepton alignment,
the bound from electron muon conversion can be relaxed to below
TeV. Note that the unavoidable bound from lepton universality in pion
decay, $M_S\gsim3.4\ TeV$, holds in all models \Leurer.

We now present an explicit example. Let the horizontal symmetry
group be $\H=U(1)_{H_1}\times U(1)_{H_2}$, with breaking parameters
$\l_1\sim\l$ and $\l_2\sim\l^2$. Assume the following charge
assignments:
\eqn\LQcharge{\matrix{
L_1&L_2&L_3&&\bar\ell_1&\bar\ell_2&\bar\ell_3&&\phi_u&\phi_d\cr
(0,2)&(2,0)&(0,0)&&(6,-1)&(-1,2)&(1,1)&&(0,0)&(0,0)\cr
Q_1&Q_2&Q_3&&\bar d_1&\bar d_2&\bar d_3&&S&\SP\cr
(3,0)&(0,1)&(0,0)&&(-1,2)&(4,-1)&(0,1)&&(-3,-2)&(3,2)\cr
&&&&\bar u_1&\bar u_2&\bar u_3&&&\cr
&&&&(-1,2)&(1,0)&(0,0)&&&\cr}}
It leads to the required hierarchy in quark and lepton parameters
and to the following leptoquark couplings:
\eqn\LQlsaqsa{
G^{\ell u}_L\sim\pmatrix{1&\l&\l^3\cr
\l^6&\l^7&\l^9\cr \l^4&\l^5&\l^7\cr},\ \ \
G^{\nu d}_L\sim\pmatrix{1&\l^5&\l^3\cr
\l^2&\l^7&\l^5\cr \l^4&\l^9&\l^7\cr}.}
Note in particular that $(G^{\ell u}_L)_{21}\sim\l^6$
so that \Glubound\ requires $M_S\gsim600\ GeV$, while
$(G^{\nu d}_L)_{12}\sim\l^5$
so that \Gndbound\ requires $M_S\gsim10\ GeV$. In this model,
the strongest bound comes from $D-\bar D$ mixing,
\eqn\LQDmix{M_S\gsim20\ TeV\times(G^{\ell u}_L)_{12}\sim5\ TeV.}

We conclude with two comments:
\par $(i)$ The horizontal symmetry may allow, in similar ways
to the models above, light leptoquarks that couple dominantly
to second or third generation fermions. While these will not
be directly produced in existing machines, they may affect
precision measurements at the $Z^0$ pole
\ref\meg{G. Bhattacharyya, J. Ellis and K. Srdihar, Phys. Lett. B336
(1994) 100, (E) B338 (1994) 522; \hfill\break
J.K. Mizukoshi, O.J.P. Eboli and M.C. Gonzalez-Garcia,
CERN preprint CERN-TH-7508-94, hep-ph/9411392.}.
\par $(ii)$ The horizontal symmetry may allow light leptoquarks
that couple dominantly to right-handed fermions, {\it i.e.}
a light $\SP$. In this case, all bounds from FCNC can be
avoided, and only the unavoidable bound from atomic parity
violation \Leurer, $M_{\SP}\gsim2\ TeV$, holds.

\newsec{Exact Horizontal Symmetries}
If there is an exact horizontal symmetry in Nature, it implies
either degeneracy between fermions or vanishing mixing angles \Gatt\lnsa.
In the quark sector, it is experimentally known that no two
quarks are degenerate and that none of the CKM angles vanishes.
Consequently, any horizontal symmetry acting non-trivially on
quarks, has to be broken. However, as lepton mixing angles have
not been experimentally determined, the possibility of an unbroken
horizontal symmetry acting on the leptons remains open.
In this section we investigate this possibility.

If we attribute to the breaking parameter $\l$ a unit $H$-charge,
then a sub-group of the horizontal symmetry remains unbroken if
there exist lepton fields with fractional $H$-charges. The selection
rule \Omixing\ is modified into
\eqn\ubmixing{\sin\theta_{ij}=0\ {\rm for}\
H(L_i)-H(L_j)={\rm non-integer}.}
As an example, we take the case where $H(L_1)$ and $H(L_2)$ are
half-integers while $H(L_3)$ is integer.
Requiring that all charged lepton masses are non-vanishing, this
would automatically require that $H(\bl_1)$ and $H(\bl_2)$ are
half-integers while $H(\bl_3)$ is integer. This can be easily
modified to any other case. The three relations \Orelations\ are
modified into
\eqn\ubrelations{
{m_\ne\over m_{\nm}}\sim\sin^2\theta_{12},\ \ \
\sin\theta_{23}=\sin\theta_{13}=0.}
Note that the last two predictions, $\sin\theta_{i3}=0$, are
not order of magnitude estimates but {\it exact} predictions:
they result from an unbroken $Z_2$ symmetry, tau-parity ($R_\tau$). Under
$R_\tau$, $L_3\rightarrow-L_3$, $\bl_3\rightarrow-\bl_3$,
while all other fields are $R_\tau$-even. Such symmetry has
interesting phenomenological implications. It forbids
$\Delta\tau=1$ processes such as $\tau\rightarrow\ell\gamma$ and
$\tau\rightarrow\ell\ell\ell$ (where $\ell$ stands here for
$e$ or $\mu$). In particular, $\nt-\nm$ and $\nt-\ne$ oscillations
are forbidden, in which case there will be no signal in the
CHORUS, NOMAD and E803 experiments. As $\nt$ is stable in this case,
cosmological constraints imply that it should be lighter
than $\O(100\ eV)$.

Note that $R_\tau$ could be an exact symmetry even if we adopt
the phenomenological input \darkmatter\ and \solar.
The hierarchy in Eqs. \lcharged\ and \lneutral\ with
\eqn\ublneutral{m_\nm/m_\nt\sim\l^5}
determines a unique set of $H$ charge assignments:
\eqn\fivecharge{\matrix{
L_1&L_2&L_3&&\bar\ell_1&\bar\ell_2&\bar\ell_3&&\phi_u&\phi_d\cr
(9/2)&(5/2)&(0)&&(7/2)&(5/2)&(3)&&(0)&(0).\cr}}
The lepton mass matrices have then the following form:
\eqn\fivemasses{
M^\ell\sim\vev{\phi_d}\pmatrix{\l^8&\l^7&0\cr\l^6&\l^5&0\cr0&0&\l^3\cr},
\ \ \ M^\nu\sim{\vev{\phi_u}^2\over\tM}\pmatrix{\l^9&\l^7&0\cr
\l^7&\l^5&0\cr 0&0&1\cr}.}
It predicts
\eqn\fivepredict{
\sin\theta_{23}=0,\ \ \ \sin\theta_{13}=0,\ \ \ m_\ne/m_\nm\sim\l^4.}
Both the form of the mass matrices \fivemasses\ and the predicted
values of the mixing angles \fivepredict\ clearly show the
consequences of tau-parity.

\newsec{Conclusions}
We have investigated the implications of supersymmetric Abelian
horizontal symmetries on the lepton sector. We find a number of
interesting predictions that hold in a large class of models:
\item{(i)} Mass ratios in the neutrino sector are of the order
of the square of the corresponding mixing angles.
\item{(ii)} The mixing angles are larger than the corresponding
ratios of charged lepton masses.
\item{(iii)} Mass ratios among neutrinos are larger than the square
of the corresponding charged leptons mass ratios.
\item{(iv)} There is no inverted hierarchy among neutrinos.

The result (iii), when combined with cosmological constraints,
implies that it is very likely that $m_\ne$, $m_\nm$ and $m_\nt$
are all lighter than $\O(100\ eV)$ \HaNi.

The embedding of the low energy selection rules in a full high
energy theory, requires the existence of new particles. However,
if the gauge symmetry does not change up to high scales, then
these particles are very likely to be too heavy to be directly
observed in experiments.

In addition to the various relations among
lepton parameters, evidence for the horizontal symmetry framework
may arise indirectly, from its implications on various other
sectors of the theory. In particular, the horizontal symmetry
suppresses SUSY contributions to lepton flavor violating
processes to an acceptable level even if sleptons are
not degenerate. If non-degenerate sleptons are discovered in
experiment, then $\mu\ra e\gamma$ is likely to be close to the
experimental upper bound.

If light leptoquarks are discovered in experiment, then again
the only likely mechanism to suppress their contributions to
FCNC processes is the horizontal symmetry. $D-\bar D$ mixing
is then likely to be close to the experimental upper bound.

\leftline{\bf acknowledgments}
We thank Ernest Baver, Michael Dine, Enrico Nardi and
Nati Seiberg for useful discussions. YN is supported in part
by the Israel Commission for Basic Research, by the
United States -- Israel Binational Science Foundation (BSF)
and by the Minerva Foundation.

\listrefs
\end